\documentclass[a4paper]{new_tlp}
\pdfoutput=1

\usepackage{mytlpnojournal}

\usepackage{xspace}
\usepackage{latexsym}
\usepackage{amssymb}   %

\usepackage{ifthen}
\newboolean{commentsaon}         
\setboolean{commentsaon}{true}
  \setboolean{commentsaon}{false}  

\newboolean{commentson} 
\setboolean{commentson}{true}

\newcommand{\comment}[1]
{\ifthenelse{\boolean{commentson}\AND\boolean{commentsaon}}
   {{\par\noindent\mbox{}{\small\footnotesize\blue[ *** #1 ]\par}\noindent\par}}
   {}}

\newcommand{\commenta}[1]
{\ifthenelse{\boolean{commentsaon}}
   {{\par\noindent\mbox{}{\small\color[rgb]{0, .5, 0}[ *** #1 ]\par}\noindent\par}}{}}
\usepackage[yyyymmdd]{datetime}

\renewcommand{\today}{2022-02-07}

\newcommand{\myhalfmagnification}{0}

\usepackage{color}
\newcommand\blue     {\color{blue}}

\newcommand{\dblue}{\color[rgb]{0,0,.9}}

\usepackage{hyperref}

\newtheorem{example}{Example}

\newcommand*{\seq}[2][n]  {{#2_{1}, \allowbreak \ldots, \allowbreak #2_{#1}}}

\newcommand*{\mydash}{{\mbox{\tt-}}}
\newcommand*{\myunderscore}{\mbox{\tt\symbol{95}}}

\title
    {Implementing backjumping by {\tt throw}/1 and {\tt catch}/3 of Prolog}
\author{W{\l}odzimierz Drabent \quad\today%
        {\ifthenelse{\boolean{commentson}\AND\boolean{commentsaon}}
           {\blue\quad  [Version 5.1 with private comments]}{}%
        }\\
        \small
      \begin{tabular}{c} \\[-1ex]
           Institute of Computer Science,
           Polish Academy of Sciences,\\
          and \\
           Department of Computer and Information Science,
           Link\"oping University, Sweden
          \\[.5ex]
 \mbox{{\tt drabent\,{\it at}\/\,ipipan\,{\it dot}\/\,waw\,{\it dot}\/\,pl}}
      \end{tabular}
}

\submitted{\today}
\begin{document}

\maketitle

\begin{abstract}
We discuss how to implement backjumping (or intelligent backtracking) in Prolog
programs
by means of exception handling.
This seems impossible in a general case.  
We provide two solutions.  One works for binary programs; in a general case it
imposes a restriction on where backjumping may originate.
The other restricts a class of backjump targets.
We also show how to simulate backjumping by means of backtracking and the
Prolog database.
\end{abstract}

\begin{keywords}
 Prolog, intelligent backtracking, backjumping, exception handling
\end{keywords}

\medskip\medskip
\noindent
In this note
we first explain the incompatibility between 
backjumping and the exception handling of Prolog.
Then we discuss how to employ the Prolog exception handling mechanism to
implement backjumping for definite clause programs
(Section \ref{sec.implementing}).
We present two approaches
of adding backjumping to a Prolog program.
The first approach is applicable to a restricted but broad class of cases,
including binary programs with arbitrary backjumping.  
The restriction is on from where the backjumping may originate.
In the second approach the class of available backjump targets is restricted,
so the resulting backjumping may only be an approximation of that intended.
Section \ref{sec.examples} presents an example of each approach.
The next section discusses backjumping by means of backtracking and the Prolog
database.
The report is completed by a brief discussion of the related work and conclusions.

\enlargethispage{1ex}
\section{Backjumping and Prolog exception handling}
\label{sec.backjumping}
A Prolog computation can be seen as a depth-first left-to-right traversal of an
SLD-tree (LD-tree, when no delay mechanisms are employed).  
Each node with $i$ children is visited $i+1$ times.  Moving from a node to its
parent is called backtracking.  By backjumping we mean skipping a part of
the traversal, by moving immediately from a node to one of its non-immediate
ancestors. 
Intelligent backtracking \cite{BruynoogheP84}
is backjumping in which it is known that there are no successes in the
omitted part of the SLD-tree.
(More generally, that there are
 no successes with answers distinct from those already
obtained.) 
Prolog provides an exception handling mechanism, consisting of built-in
predicates throw/1 and catch/3.  
Let us follow the Prolog standard \cite{Prolog.standard96}
and explain them in terms of LD-trees.
Let $A_c={\it catch(Q,s,Handler)}$, where $Q$ and ${\it Handler}$ are queries,
and $s$ is a term.  A node $A_c,N$ of an LD-tree
has a single child $Q,N$.  A second child
may however be created as a result of exception handling.
An exception is raised by
invoking ${\it throw(t)}$ (formally, by visiting a node
 $N_t={\it throw(t),N'}$; such node has no children).
This is sometimes called ``throwing a ball $t$''.
Visiting $N_t$ starts a search along the
path from $N_t$ to the root.  The search is for a node
$N_c={\it catch(Q,s,Handler)},N$ with one child
 such that (a) a freshly renamed copy $t'$
of the ball $t$ is unifiable with $s$, with an mgu $\theta$, and
(b) ``the ball is thrown during the execution of'' $Q$
\cite{Prolog.standard96}, in other words --
no node between $N_c$ and $N_t$ is an instance of $N$.
The first (closest to $N_t$) such node $N_c$ on the path is chosen, and
 a new child ${\it (Handler,N)\theta}$ of $N_c$ is added to the tree.
The new child becomes the next visited node of the tree.

Prolog does not provide any way to directly implement backjumping.
It may seem that exception handling
is a suitable tool for this task.  There is however an important difference.
Not all backtrack points can be reached by means of catch/3.

Consider an LD-tree containing a node $Q$ with $k$ children $Q_1,\ldots,Q_k$.
We may say that $k$ backtrack points correspond to $Q$.  Any of them may be a
target for backjumping, but exception handling is able to arrive only at
the last one.
In particular, for intelligent backtracking it may be
necessary that backjumping to $Q$
  from a descendant of $Q_i$ is followed by visiting 
  $Q_{i+1},\ldots,Q_{k}\!$ and their descendants.
Exception handling would however omit all such nodes.
More precisely, if     %
$N=catch(Q,s,H)$ is employed to catch an exception,
then $N$ has a child $Q$, and all its descendants are as above.
However catching an exception at $N$ results 
in omitting all the unexplored descendants of $Q$.
The same happens if $N=catch(Q',s,H),Q''$, where $Q=Q',Q''$.
Also, it seems that the omitted part of the tree cannot be explored
by reconstructing it
by the exception handler $H$,
at least in the general case.

A possible solution could be backjumping
to the last backtrack point of $Q_i$, instead of the $i$-th backtrack point
of $Q$.
(This idea is exploited in Approach 1 below.)
However, to implement such a backjump by catch/3 in a general case, one requires
replacing $Q_i$ by ${\it catch(Q_i,\ldots)}$ (as the backjump may come from
any descendant of $Q_i$).  In many cases there is no way of adding catch/3 to
the original program to obtain such a query.
\footnote{
  Consider $Q=A,B$ and $Q_i=(B_i,B)\theta$; assume that clause $A'\gets B_i$
  was used in the resolution step.  Adding ${\it catch}$ to the clause
  results in $B\theta$ not being a part of the argument of ${\it catch}$.
  Adding  ${\it catch}$ elsewhere in the program results in backjumping to
  another node of the tree.
}

This discussion shows that backjumping cannot be, in general, directly
implemented by means of Prolog exception handling.
It refutes the claim contained in the title of a recent paper
\cite{DBLP:journals/tplp/RobbinsKH21},
 which says ``backjumping is exception handling''.%
\footnote{
  We should also mention differences not related to implementing backjumping.
  In exception handling,  after an exception is caught, the exception
    handler
  is activated.  In backjumping
    there is nothing similar to an exception handler.
  Also, in contrast to backjumping, exception handling makes it possible to
  pass information (an arbitrary term) from the point where the exception is
  raised to the one where it is caught.  This is done by means of the
  argument of  ${\it throw}$/1.
}

\enlargethispage{.5ex}

\section{Implementing backjumping by exception handling}
\label{sec.implementing}
\subsection{Approach 1}
\label{sec.approach1}
Now we discuss a way of implementing backjumping by employing Prolog
exception handling.
Assume that we deal with a definite clause program $P$, which we want to
execute with backjumping.
The target of backjumping is to be identified by a term $id$.  
So backjumping is initiated by $throw(id)$.

Assume that the
target of backjumping is a node $A,Q$ of the LD-tree, where $A$ is an atom.
Assume that $A,Q$ has $k$ children, $\seq[k]Q$.
Let $p$ be the predicate symbol of $A$ and 
\begin{equation}
\label{theprogram}
\begin{array}{l}
  p(\vec t_1)\gets B_1. \\ \cdots \\     p(\vec t_n)\gets B_n.
\end{array}
\end{equation}
where $k\leq n$,
be the procedure $p$ of program $P$
(i.e.\ the clauses of $P$ beginning with $p$).

Consider backjumping initiated by $throw(id)$ in the subtree rooted in $Q_i$.  
The subtree should be abandoned, but the descendants of $Q_{i+1},\ldots,Q_k$
should not. Thus we need to restrict the exception handling to this subtree.
A way to do this is to replace each $B_j$ by
$
catch( B_j, id, {\it fail} )
$.
Then performing $throw(id)$ while executing $B_j$ results in failure of the 
clause body and backtracking to the next child of $A,Q$, as required.
Assume that a query $b t id(\vec t,Id)$ 
\newcommand{\myunderline}[1]{\makebox[0pt][l]{\underline{#1}}}%
(\myunderline{b}backjump \myunderline{t}target \myunderline{id}identifier)
 produces the unique identifier $id$
out of the arguments of $p$.  
 The backjumping is implemented by a transformed procedure consisting of
clauses
\begin{equation}
\label{eq.approach1.1}
{\it
p(\vec t_j)\gets b t id(\vec t_j,Id), \,
  catch( B_j, Id,  fail ) }.
\qquad\qquad \mbox{for }j=1,\ldots,n
\end{equation}
(where ${\it Id}$ is a variable).

Transforming a program in this way correctly implements backjumping, however
with an important limitation.
Speaking informally,
backjumping to a node with $p(\vec t)$ selected must occur while executing 
$p(\vec t)$.
Otherwise the exception is not caught and the whole computation is abandoned.

An important class of programs which satisfy this limitation are binary logic
programs (i.e.\ programs with at most one body atom in a clause).  
The approach presented here works for
such programs and arbitrary backjumping.

Sometimes (like in Ex.\,\ref{ex.binary} below)
it may be determined in advance that, for some $j$, no exception
will be caught by the ${\it catch}$/3 in  (\ref{eq.approach1.1}).
So in practice
some clauses of (\ref{theprogram}) may remain unchanged
(or a choice between $B_j$ and  ${\it catch( B_j, Id,  fail)}$
may be made dynamically,
e.g.\ 
 by modifying the body of  (\ref{eq.approach1.1}) into
${\it
b t id(\vec t_j,Id) %
    \to    %
      catch( B_j, Id, {\it fail} )  %
    \linebreak[3]
\mathop;
 B_j 
}
$).

\paragraph{Approach 1a.}
\label{proba.paragraph}
Here we present a variant of Approach 1.
Roughly speaking, in the former approach
control is transferred to the next clause due to failure of a clause body.
So catching an exception causes an explicit failure.
Here control is transferred to the next clause by means of an exception,
so standard backtracking eventually raises an exception.
To simplify the presentation we assume that in  (\ref{theprogram}) all
the clause heads are the same, 
$\vec t_1=\cdots=\vec t_n=\vec t$.

Assume first that $n=2$.  Backjumping equivalent to that of Approach 1 can be
implemented by
\begin{equation}
\label{program.approach1asmall}
\begin{array}{l}
\it
p(\vec t) \gets  b t id(\vec t,Id),
     catch(
     \begin{array}[t]{l}
\it      (B_1 \mathrel; throw(Id)),   \\     
\it       Id,  \\
{\it       catch( B_2, Id, {\it fail})\ ).}
     \end{array}
\end{array}
\end{equation}
Invocation of $B_2$ is placed in the exception handler,
so we additionally raise an exception when $B_1$ (the first clause body) fails.
For arbitrary $n$, the transformed procedure (\ref{theprogram}) is:
\vspace{-1ex plus 1 ex}
\begin{equation}
\label{program.approach1a}
\begin{minipage}{.9\textwidth}
\vspace*{-1\abovedisplayskip}
 \[
    p(\vec t) \gets
    \begin{array}[t]{l}
\it     b t id(\vec t,Id), \\
         catch(
         \begin{array}[t]{l} (\it B_1 \mathrel; throw(Id)), \\
\it                    Id ,              \\
         catch(
         \begin{array}[t]{l} (B_2 \mathrel; throw(Id)), \\
\it                    Id, \\ \ldots \\
         catch(
         \begin{array}[t]{l} (B_{n-1} \mathrel; throw(Id)), \\
\it                    Id ,              \\
\it         catch(
                    B_{n},\,  %
                    Id ,\,             %
                    {\it fail} \,)\ \, )\cdots )).
    \end{array}\end{array}\end{array}\end{array}
    \hspace*{-3em}
\]
\end{minipage}%
\end{equation}

Generalizing this transformation to clauses with different heads
is rather obvious.  The same for 
employing a different backjump target identifier for each clause.
 Note that 
in this approach it is possible to augment backjumping by passing information
(from the place where backjump originates to the backjump target).%
\footnote{
To pass a term $t$, one may choose the backjump target identifier to be 
$f(X_i)$ for clause $i$.  Then performing ${\it throw}(f(t))$ while executing
$B_i$ results in binding $X_i$ to $t$ when the exception is caught.
This makes $t$ available in those bodies $B_{i+1},\ldots,B_n$ that contain
$X_i$.  
E.g.\ for $n=2$
instead of the body of (\ref{program.approach1asmall}) we obtain
${\it
catch(\, (B_1 ; throw(f(no b j))),\, f(X_1),\, 
    catch(B_2,   f(X_2), fail) \, )
}$;  constant ${\it no b j}$ (for ``no backjumping'') is passed when 
standard backtracking takes place.
\vspace{.5ex}
} %
Such augmenting is impossible in Approach 1 and Approach 2 below.

\subsection{Approach 2, approximate backjumping}
\newcommand{\myfigBB}{%
    \begin{minipage}[t]{.35\textwidth}
\[
      \begin{array}[t]{c}
        N_0=A_0,Q_0 \\[-1ex]  \rule{.5pt}{1ex} \\[-1ex]
        N_1=(B_0,B_1,Q_0)\theta\\ \vdots \\
        {\dblue N}=A,\!Q = Q_2,\!(B_1,Q_0)\theta\rho
        \\  \vdots \\
       N_2=A_2,\ldots,(B_1,Q_0)\varphi'    \\  \vdots \\
      N''=(B_1,Q_0)\varphi     \\  \vdots \\
        {\dblue N'}=A',\ldots,Q_0\psi\\  %
      \end{array}
\]
    \end{minipage}
}

\noindent
We have shown how to implement backjumping to an LD-tree node $N = A,Q$
(with atomic $A$)
from within the execution of $A$.
(Formally: no node between $A,Q$ and the origin of backjumping is an
instance of $Q$.)
It remains to discuss backjumping originating in the execution of $Q$.
Assume that the initial query is atomic; dealing with arbitrary initial
queries is similar.
In such case, the program contains a clause $H{\gets}B_0,B_1$ %
(where $B_0,\,B_1$ are nonempty), %
such that, speaking informally,
 the backjumping is from within the execution of $B_1$,
and its target is within the execution of $B_0$.%
\footnote{\label{bigfootnote}
  Let us provide a detailed explanation.
  Note first that 
  for any (occurrence of an) atom $A$ in a node $N$ of an LD-tree,
  and any ancestor $N_0$ of $N$, there exists a unique (occurrence of an)
  atom $A_0$ in $N_0$, such that $A$ has been derived from $A_0$ in the
  resolution steps between $N_0$ and $N_1$.  
  We omit a detailed formalization of this correspondence.
  Let us denote such occurrence (of $A_0$ in $N_0$) by ${\it pre}(A,N,N_0)$.
\vspace{.5ex}

 \begin{minipage}[t]{.65\textwidth}
  Assume that node $N=A,Q$ is the target of backjumping, and that its origin is
  $N'$.  Let $A'$ be the first atom of $N'$.
  Consider the closest ancestor $N_0$ of
  $N$ such that ${\it pre}(A,N,N_0)={\it pre}(A',N',N_0)$.
  Let $A_0={\it pre}(A',N',N_0)$.
  Consider the child $N_1$ of $N_0$ that is an ancestor of $N$.
  So  $N_0=A_0,Q_0$, and 
  $N_1=(B_0,B_1,Q_0)\theta$,
  where ${\it pre}(A,N,N_1)$ occurs in $B_0\theta$, and 
  ${\it pre}(A',N',N_1)$ in  $B_1\theta$.
  Moreover $N_1$  was obtained by resolving $N_0$ with a clause
  $H{\gets}B_0,B_1$.  (Note that splitting its body into $B_0$ and $B_1$ may
  be not unique.) 

\vspace{.5ex}

  Now $N$ can be represented as $N=Q_2,(B_1,Q_0)\theta\rho$.
  No node between $N_1$ and $N$ is an instance of $B_1,Q_0$.
 So, informally, the backjump target $N$ is within the execution of $B_0\theta$.

\vspace{.5ex}
  Note that if a descendant of $N$ is of the form
  $N_2=A_2,\ldots,(B_1,Q_0)\varphi$ then  
  ${\it pre}(A_2,N_2,N_1)$ occurs in  $B_0\theta$.
  Thus $N$ has a descendant of the form $N''=(B_1,Q_0)\varphi$ 
  (otherwise $N'=A',\ldots$ is of the same form as $N_2$, thus
  ${\it pre}(A',N',N_1)$ is in  $B_0\theta$, contradiction).
  As ${\it pre}(A',N',N_1)$ occurs in  $B_1\theta$, 
  no node between $N_1$ and $N'$ (including $N'$) is an instance of $Q_0$.
  Hence $N'$ is of the form $A',\ldots,Q_0\psi$,
  and thus the backjump origin $N'$ is within the execution of $B_1\theta$.
 \end{minipage}%
 \myfigBB
} 

\pagebreak[3]
Such backjumping exactly to the target does not seem possible to be
implemented by means of {\it throw}/1 and {\it catch}/3.
However we may force $B_1$ to fail when an
exception is thrown.
This means backjumping to, speaking informally, the
success of $B_0$, instead of the original target $N$.  
(In the notation of footnote \ref{bigfootnote}, 
the target of this backjump is $N''=B_1\varphi,\ldots$.)
This in a sense approximates backjumping to $N$.  
In some cases such shorter backjumping may still be useful.
It may exclude from the search space a major part of what would
be excluded by backjumping to $N$.

To implement such approximated backjumping
we need to change the program, so that the instance $B_1\varphi$ of $B_1$
in node $N''$ is replaced by  $catch(B_1\varphi,id,{\it fail})$.  
To obtain this, the clause
\[
H\gets B_0,B_1
\ \quad \mbox{ is transformed to } \ \quad
    H\gets B_0,\, b t id(\ldots,Id), \,
    catch(B_1,Id,{\it fail})
\]
where $b t id$, as previously, is used to obtain the unique identifier for
the backjump target.

\color{black}\normalsize %

\newcommand\tttrue{\mbox{\tt true-}}%
\newcommand\ttfalse{\mbox{\tt false-}}%

\section{Examples}
\label{sec.examples}
We apply the approaches introduced above to a simple program,
a naive SAT solver.  It uses the representation
of clauses proposed by  \citeN{howe.king.tcs-shorter}.
(Note that we deal here with two kinds of clauses -- those of the program, and
the propositional clauses of a SAT problem.)
A conjunction of clauses 
is represented as a list of (the representations of) clauses.
A clause is represented as a list of (the representations of) literals.
A positive literal is represented as a pair
$\tttrue X$ and a negative one as $\ttfalse X$, where the Prolog variable
represents a propositional variable.
For instance a formula
$(x\lor\neg y\lor z)\land(\neg x\lor v)$ is represented as 
{\tt[[true-X,false-Y,true-Z],[false-X,true-V]]}.
In what follows we do not distinguish literals, clauses, etc from their
   representations.

Thus solving a SAT problem for a conjunction of clauses $sat$
means instantiating the variables of $sat$ in such way that each of the lists 
contains an element of the form $t\mydash t$.
This can be done by a program $P_1$:  %
      \[
      \begin{array}{l}
          sat\_cl([Pol\mydash {\it Pol} | {\it Pairs}]). 
      \label{satcl1}
    \\
              sat\_cl([  H            | {\it Pairs}]) \gets 
                sat\_cl({\it Pairs}).
                \\
                {\it sat\_c n f}([\,]). \label{satcnf1}  \\
    {\it sat\_c n f}([Clause | Clauses]) \gets
        sat\_cl(Clause),
        \ {\it sat\_c n f}(Clauses). \label{satcnf2}  
      \end{array}
      \]
and a query     ${\it sat\_c n f}(sat)$.
See \cite[Section 3]{Drabent.tplp18} 
for further discussion and
a formal treatment of the program.

We add backjumping to $P_1$.  
The intention is that, 
after a failure of ${\it sat\_c l}(cl)$ (where $cl$ is the representation of a
partly instantiated clause) a backjump is performed
to the last point where a variable from clause $cl$ was assigned a
value.  
This does not correctly implement intelligent backtracking,%
\footnote{
  E.g. for $(x\lor y)\land(\neg z\lor z)\land(\neg x\lor\neg y)\land
  (\neg x\lor y\lor z)$ no solution with $z$ being \mbox{\tt true} is found.
  An explanation is that, speaking informally, backjumping from the last clause
  (with $x,y,z$ instantiated to
  {\tt true}, {\tt false}, {\tt false})
  arrives to the previous one
  (where $y$ was set to false), this immediately causes backjumping to the
  first clause.
  }  %
 but the purpose
is to illustrate the approaches proposed in the previous section.

\pagebreak
\begin{example}
\label{ex.approach2}
  Here we employ Approach 2 to program $P_1$.  Speaking informally, the
  required backjumping originates from within
 ${\it sat\_c n f(Clauses)}$  in the last clause of the program,
and its target is in     ${\it sat\_cl(Clause)}$.
 We approximate this backjumping by a failure of  ${\it sat\_c n f}$.
    (Note that in this case the approximation is good, 
    the intended target is a node of the form
     ${\it sat\_cl}([v\mydash V|t]),{\it sat\_c n f}(t')$
    and we implement backjumping to its child 
     ${\it sat\_c n f}(t'\{V/v\})$.)

We augment the values of variables; %
   the value of a variable is going to be of the form
 $(l,v)$, where $l$ is a number (the {\em level} of the variable)
   and $ v$ a logical value {\tt true} or {\tt false}.
The level shows at which recursion depth of ${\it sat\_c n f}$ the value was
assigned.  The levels will be used as identifiers for backjump targets. 
In such setting,
a substitution $\theta$ assigning values to variables makes a SAT
 problem $sat$ 
 satisfied when each member of list
 $sat\theta$ contains a pair
 of the form $ v\mydash(l, v)$.
 This leads to transforming the first clause of $P_1$ to 
 ${\it sat\_cl([Pol\mydash (\myunderscore,{\it Pol}) | {\it Pairs}])}$.

We transform $P_1$ into a program $P_2$ which takes levels into account.
We add the current level as the second argument
of ${\it sat\_c n f}$ and of ${\it sat\_c l}$, and we add a third argument
to  ${\it sat\_c l}$.
The declarative semantics of the new program is similar to that of $P_1$;
the answers of $P_2$ are as follows.
  If
  the first argument of ${\it sat\_c l}$ is a list then it 
  has a member of the form $t\mydash(t',t)$.  
  Also, this condition is satisfied by each element of the list that is 
  the first argument of ${\it sat\_c n f}$.

  Operationally, an invariant will be maintained that, whenever 
  ${\it sat\_cl(cl,l,h l)}$ is selected in LD-resolution, $cl$ is a list and
  $l$ and $h l$ are numbers, 
  $l>h l$ and $l$ is greater than any number occurring in $cl$.
  List $cl$ is the not yet processed fragment of a clause $cl_0$
(possibly instantiated),
$l$ is the current level, and
${\it h l}$ is the highest level of those variables that occur in the already
processed part of $cl_0$ and
have been bound to some values at previous levels;
${\it h l}=-1$ when there is no such variable.
  In case of failure of ${\it sat\_c l(cl,l,h l)}$,
  an exception will be raised with the ball being the maximum of $h l$
  and the levels of the variables occurring in $cl$ (provided the maximum is
  $\geq0$).

Checking the value already assigned to a variable must be treated differently
  from assigning a value to an unbound variable.
This leads to two clauses playing the role of the first clause of $P_1$.
So procedure  ${\it sat\_c l}$ of  $P_1$ is transformed into the following
procedure of $P_2$:
\vspace{0pt minus 0.6ex}
\[
\begin{minipage}[t]{.75\textwidth}
\begin{verbatim}
sat_cl( [Pol-V|_Pairs], _L, _HL ) :-
        nonvar(V), V=(_,Pol).
sat_cl( [Pol-V|_Pairs], L, _HL ) :-
        var(V), V=(L,Pol).
sat_cl( [_-V|Pairs], L, HL ) :-
        new_highest( V, HL, HLnew ),
        sat_cl( Pairs, L, HLnew ). 
\end{verbatim}
\end{minipage}
\makebox[0pt][l]{\hspace{3.3em}
    \begin{minipage}[t]{.05\textwidth}
        (\refstepcounter{equation}\theequation\label{clause1.ex.P2})
\\\\        (\refstepcounter{equation}\theequation)
    \label{clause2.ex.P2}
\\\\\\        (\refstepcounter{equation}\theequation)
    \label{clause3.ex.P2}
    \end{minipage}
} %
\vspace{0.6ex}
\]
%
%
Predicate ${\it new\_highest}$ takes care of updating the highest level
of the variables from the already processed part of the clause.
\[
   \begin{minipage}[b]{.76\textwidth}  %
{\small%
  \begin{oldtabular}{@{}l}
    \% {\tt  new\_highest}${\it(var, h, h new)}$ -- if
       ${\it var}$ is a Prolog variable
    then $h =h new$ \\
    \% \qquad\qquad\ otherwise ${\it var=(l, v)}$ and $h new =\max(h,l)$
    \vspace{-1ex}
  \end{oldtabular}
}
\begin{verbatim}
new_highest( V, H, H ) :- var( V ).
new_highest( V, H, H ) :- nonvar( V ), V=(L,_Value), H>=L.
new_highest( V, H, L ) :- nonvar( V ), V=(L,_Value), H<L.
\end{verbatim}
  \end{minipage}
\makebox[0pt][l]{\hspace{2.6em}%
    \begin{minipage}[b]{.05\textwidth}  %
\hfill   %
 (\refstepcounter{equation}\theequation\label{clause1aux.ex.P2})
\\ \mbox{}\hfill  
 (\refstepcounter{equation}\theequation\label{clause2aux.ex.P2})
\\ \mbox{}\hfill  
 (\refstepcounter{equation}\theequation\label{clause3aux.ex.P2})
    \end{minipage}
}  %
\]
Procedure ${\it sat\_c n f}$ is transformed into
\vspace{-.9ex}
\[
\begin{minipage}[t]{.46\textwidth}
\begin{verbatim}
sat_cnf( [], _L ).
sat_cnf( [Clause|Clauses], L ) :-
        sat_cl( Clause, L, -1 ),
        Lnew is L+1,
        sat_cnf( Clauses, Lnew ).
\end{verbatim}
\end{minipage}
\makebox[0pt][l]{\hspace{8.6em}%
    \begin{minipage}[t]{.05\textwidth}
\hfill   %
 (\refstepcounter{equation}\theequation\label{clause7.ex.P2})
\\\\[1.5ex] \mbox{}\hfill  
 (\refstepcounter{equation}\theequation\label{clause8.ex.P2})
    \end{minipage}
}  %
\]
\newcommand*{\myindent}{\hspace{5em}}%
Program $P_2$ consists of clauses (\ref{clause1.ex.P2}) -- (\ref{clause8.ex.P2}).
An initial query
${\it sat\_c n f}(sat,0)$ results in checking the
satisfiability of a conjunction of clauses $sat$.
Now we add backjumping to $P_2$.
The backjumping has to be triggered instead of a failure of ${\it sat\_cl}$.
The latter happens when the first argument of  ${\it sat\_cl}$ is $[\,]$.
The new program $P_3$ contains the procedure ${\it sat\_cl}$ of $P_2$, and
additionally a clause
\begin{equation}
\label{clause.throw.ex.P3}
\begin{minipage}[t]{.74\textwidth}
\begin{verbatim}
sat_cl( [], _, HL ) :- HL>=0, throw( HL ).
\end{verbatim}
\end{minipage}
\end{equation}
triggering a backjump.
When ${\it H L}<0$  then there is no target for backjumping, 
and standard backtracking is performed.

The procedure ${\it sat\_c n f}$ of the new program $P_3$, is constructed out
of that of $P_2$ by transforming clause (\ref{clause8.ex.P2})
as described in Approach 2:
\begin{equation}
\label{clause2.ex.P3}
\begin{minipage}{.53\textwidth}
\begin{verbatim}
sat_cnf( [Clause|Clauses], L ) :-
        sat_cl( Clause, L, -1 ),
        Lnew is L+1,
        catch( sat_cnf( Clauses, Lnew ),
               L,
               fail
              ).
\end{verbatim}
\end{minipage}
\end{equation}
So backjumping related to the variable with level $l$, implemented as 
$throw(l)$, arrives to an instance of clause (\ref{clause2.ex.P3})
 where %
$L$ is $l$.  The whole $catch(\ldots)$
fails, and the control backtracks to
the invocation of ${\it sat\_c l}$ that assigned the variable.
(An additional predicate ${\it b t id}$ was not needed, as
$L$ is the unique identifier.)
Now program $P_3$ consists of clauses
(\ref{clause1.ex.P2}) -- (\ref{clause7.ex.P2}) and
(\ref{clause.throw.ex.P3}) -- (\ref{clause2.ex.P3}). 
To avoid leaving unnecessary backtrack points in some Prolog systems,
each group of clauses with ${\it var}$/1 and ${\it non var}$/1 
(clause (\ref{clause1.ex.P2}) with (\ref{clause2.ex.P2}), and 
(\ref{clause1aux.ex.P2}) with (\ref{clause2aux.ex.P2}) and
(\ref{clause3aux.ex.P2}))
may be replaced by a single
clause employing
$({\it var(V) \mathop\to \ldots{;}\ldots })$ and, in the second case,
$({\it H{<}L \mathop\to \ldots{;}\ldots })$.
To simplify a bit the initial queries, a top level predicate may be added,
defined by a clause \ \
{\small\verb|sat(Clauses) :- sat_cnf(Clauses,0).|}

\end{example}

\pagebreak[3]
\begin{example}\rm
\label{ex.binary}
    Here we transform $P_1$ from Ex.\,\ref{ex.approach2}  to a binary program and apply Approach 1.
    The binary program $P_{\mathrm b}$ is
\vspace{-1ex}
      \[
      \begin{array}{l}
                {\it sat\_b}(\,[\,]\,).   \\
    {\it sat\_b}(\,[ [{\it Pol\mydash Pol}|\myunderscore] | Clauses]\,) \gets
        \ {\it sat\_b}(Clauses). 
    \\
    {\it sat\_b}(\,[ [\myunderscore|Pairs] | Clauses]\,) \gets
    {\it sat\_b}([Pairs | Clauses]).
      \end{array}
      \]
Note that %
in Ex.\,\ref{ex.approach2} the unprocessed part of the current clause was an
argument of ${\it sat\_cl}$, now it is the head of the %
 argument of ${\it sat\_b}$.
In what follows we do not explain some details which are as 
in the previous example.
As previously we introduce levels,
and represent a value of a variable by $(l, v)$, where $l$ is a 
level and $ v$ a logical value.  
As previously, we first transform  $P_{{\rm b}}$ into 
$P_{{\rm b}2}$ dealing with levels, and then add backjumping to $P_{{\rm b}2}$.
We add two arguments to ${\it sat\_b}$,
they are the same as the arguments added to  ${\it sat\_cl}$ in
Ex.\,\ref{ex.approach2}.  
The declarative semantics is similar, the first argument of 
${\it sat\_b}$ (in an answer of $P_{{\rm b}2}$) is
 as the first argument of ${\it sat\_c n f}$ in 
 $P_2$.
An invariant similar to that of Ex.\,\ref{ex.approach2} will be maintained
by the operational semantics.
Whenever ${\it sat\_b(cl s,l,h l)}$ is selected,
$l$ and $h l$ are numbers,
$l>h l$ and $l$ is greater than any number occurring in $cl s$.
List $cl s$ is a conjunction of clauses (possibly instantiated),
and its head, say $cl$, is the not yet processed fragment of the current
clause, say $cl_0$;
number $l$ is the current level, and
${\it h l}$ is the highest level of variables from the already processed part
of $cl_0$.
Now program $P_{{\rm b}2}$ is:
\[
\begin{minipage}[t]{.75\textwidth}
\begin{verbatim}
sat_b( [], _L, _HL ).
sat_b( [[Pol-V|_] | Clauses], L, _HL ) :- nonvar(V),  
        V=(_,Pol), Lnew is L+1,
        sat_b( Clauses, Lnew, -1 ).
sat_b( [[Pol-V|_] | Clauses], L, _HL ) :- var(V), 
        V=(L,Pol), Lnew is L+1,
        sat_b( Clauses, Lnew, -1 ).
sat_b( [[_-V|Pairs] | Clauses], L, HL ) :- 
        Lnew is L+1,
        new_highest( V, HL, HLnew ),
        sat_b( [Pairs | Clauses], Lnew, HLnew ).
\end{verbatim}
\end{minipage}
\makebox[0pt][l]{\hspace{2.8em}
    \begin{minipage}[t]{.05\textwidth}
        (\refstepcounter{equation}\theequation\label{clause1})%
    \label{program.exbinary2}%
    \label{clause1.ex.binary2}%
\\\\
        (\refstepcounter{equation}\theequation)%
    \label{clause2.ex.binary2}%
\\\\\\        (\refstepcounter{equation}\theequation)%
    \label{clause3.ex.binary2}%
\\\\\\ [1ex]        (\refstepcounter{equation}\theequation)
    \label{clause4.ex.binary2}
    \end{minipage}
}
\]
Procedure ${\it new\_highest}$/3 is the same as in the previous example.
Program $P_{{\rm b}2}$ with a query ${\it sat\_b}(sat,0,-1)$
checks satisfiability of the conjunction of clauses $sat$.

Now we add backjumping to $P_{{\rm b}2}$.  As previously, backjumping originates 
when an empty clause is encountered:
\begin{equation}
\label{clause5throw.ex.binary2}
\begin{minipage}{.75\textwidth}
\begin{verbatim}
sat_b( [[] | _Clauses], _L, HL ) :-  HL>=0, throw( HL ).
\end{verbatim}
\vspace{-1\abovedisplayskip}
\end{minipage}
\end{equation}
Let us discuss backjump targets.
Assume that the nodes of an LD-tree satisfy the invariant.
Consider the descendants of a node $N = {\it sat\_b}(cl s,l,h l)$ 
 obtained by first resolving $N$ with clause (\ref{clause2.ex.binary2}) or
(\ref{clause4.ex.binary2}).
A ball thrown from such a descendant $N_2$ is not $l$.%
\footnote{
  Consider a node $N = {\it sat\_b}(cl s,l,h l)$ and its closest descendant $N'$
   of the form ${\it sat\_b(\ldots)}$.
   So $N'={\it sat\_b(\ldots,l{+}1,\ldots)}$.
   If a number $i$ occurs in a node between $N$ and $N'$, or in $N'$,
    then $i=l{+}1$ or $i$ occurs in $N$.
   By induction, if $i$ occurs in an descendant of $N$ then $i$ occurs in $N$ or
   $i>l$. 
   Additionally, if $N'$ was obtained by first resolving $N$ with
   (\ref{clause2.ex.binary2}) or (\ref{clause4.ex.binary2}), then $N'$ does not
   contain $l$.  
   Thus no descendant of $N'$ contains $l$.
}  
So for the backjump target we need to modify only clause
(\ref{clause3.ex.binary2}).  Following Section \ref{sec.approach1} we obtain:
%
%
%
%
%
\begin{equation}
\label{clause3catch.ex.binary2}
\begin{minipage}{.75\textwidth}
\begin{verbatim}
sat_b( [[Pol-V|_] | Clauses], L, _HL ) :- 
        catch(
              ( var(V), V=(L,Pol), Lnew is L+1,
                sat_b(Clauses, Lnew, -1)
               ),
               L,
               fail
              ).
\end{verbatim}
\end{minipage}
\end{equation}
The final program $P_{{\rm b}3}$  consists of clauses
(\ref{clause1.ex.binary2}), (\ref{clause2.ex.binary2}), 
(\ref{clause3catch.ex.binary2}),
(\ref{clause4.ex.binary2}),  (\ref{clause5throw.ex.binary2}), 
and (\ref{clause1aux.ex.P2}) -- (\ref{clause3aux.ex.P2}).%
\footnote{
In (\ref{clause3catch.ex.binary2}), the first atom ${\it var}(V)$
  from the body of (\ref{clause3.ex.binary2})
  can be moved outside of ${\it catch}$,  %
  transforming the body of (\ref{clause3catch.ex.binary2}) to
  ${\it var(V), catch(\ldots)}$.
  (This is because ${\it var}(V)$ is deterministic and not involved in
  backjumping.)  Now, similarly as in the previous example, 
  some backtrack points may be avoided by replacing
  clauses (\ref{clause3catch.ex.binary2}) and (\ref{clause2.ex.binary2}) 
  by a single clause with the body of the form  
  ${\it var(V)\mathop{\to}\ldots \mbox{\tt;}\ldots }$.
}  

\end{example}

\section{Another approach}

Here we discuss simulating backjumping by means of Prolog backtracking.
This requires employing the Prolog database.
An example  of such approach was presented by 
\citeN{DBLP:journals/tplp/Bruynooghe04}.
A backjump is initiated by a failure preceded 
by depositing in the Prolog database an identifier of the backjump target.
At each backtracking step, the database is queried to check if the
backjumping target is reached. If not, further backtracking is caused.
This is done by some extra code placed at the beginning of the body
of each clause involved in backjumping.
(In the presented example, there is only one such clause.)

For some programs it may be impossible, or difficult, to statically determine
the clauses involved in backjumping.  Also, the set of such clauses may
differ for various initial queries.  
In a general case, the idea of \citeN{DBLP:journals/tplp/Bruynooghe04}
can be implemented 
by converting each clause $p(\vec t)\gets B$  into
\[
p(\vec t)\gets b t id(\vec t,Id),catch(Id), B.
\]
where $b t id$/2 is as in Section \ref{sec.implementing}, and $catch$/1 is a
new predicate. 
An invariant is maintained that the database contains a backjump target only
during backjumping. 
Query
$catch(t)$ succeeds immediately, unless during backjumping.  In the latter
case it fails if $t$ is not unifiable with the backjumping target.
Otherwise it removes the target from the database
and succeeds (instantiating $t$ in the obvious way).%
\footnote{
    In SICStus, 
    it can be defined by\,
    {\tt catch(Id)\,:-\,bb\myunderscore get(target,\myunderscore)\,->\,%
    bb\myunderscore delete(target,Id)\,;\,true.%
    },\,
    and use
     {\tt bb\myunderscore put(target,$t'$),fail} %
    to cause a backjump.
}

Note that there are no restrictions 
in this approach 
on the origin/target of
backjumping, in contrast to those discussed in Section \ref{sec.implementing}.

\section{Final comments}

\paragraph{Related work.}
For the work of \citeN{DBLP:journals/tplp/Bruynooghe04}, see the previous
section.
\citeN{DBLP:journals/tplp/RobbinsKH21}  present a non-trivial example of 
using Prolog exception handling to implement backjumping.
(The main example is preceded by a simple introductory one.)
The program is a SAT solver with conflict-driven clause learning.
A learned clause determines the target of a backjump.
There is no plain backtracking.
The program keeps the learned clauses in the Prolog database, to preserve
them during backjumping.
Prolog coroutining is is employed in a fundamental way.
The program is rather complicated,
it seems impossible to view it as some initial program with added
backjumping.  To understand it one has to reason about the details of the
operational semantics.

That paper  does not propose any general way of adding backjumping to logic
programs.  The difference between backjumping and Prolog exception handling
discussed here in Section \ref{sec.backjumping}  is not noticed.%
\footnote{
    Most likely this is because the authors have not faced the limitations
    pointed out here.
    Backjumping in their program is similar to that of
    (\ref{program.approach1asmall})
    (Approach 1a for $n=2$),
     with ${\it throw(Id)}$  dropped 
     (as there is no standard backtracking), and 
     ${\it catch(B_2,Id,fail)}$ replaced by $B_2$
     (as there is no backjumping from $B_2$ with the current $Id$).
} 
We cannot agree with the claims ``backjumping is exception handling"
 and that ``{\tt catch} and {\tt throw} [provide]
exactly what is required for programming backjumping''
\cite[the title, and p.\,142-143]{DBLP:journals/tplp/RobbinsKH21}.

\paragraph{Conclusions.}

The subject of this paper is adding backjumping to logic programs.
Additionally, we briefly showed how to simulate
backjumping by means of plain backtracking and the Prolog database.

%
We discussed
the differences between backjumping and Prolog
exception handling, and proposed two approaches to implement the former by
the latter.  This seems impossible in a general case.
The first approach imposes certain restrictions on where backjumping can be
started.  The second one -- on the target of backjumping.
The restrictions seem not severe.
The first approach is applicable, among others, to binary programs with
arbitrary backjumping.  
For the second approach, 
the presented example shows that sometimes the difference between the
required and the actual target
may be unimportant.
As every program can be transformed to a binary one
\cite{DBLP:books/mk/minker88/Maher88,DBLP:conf/plilp/TarauB90},
the first approach is indirectly applicable to all cases.
\bibliographystyle{acmtrans}
\bibliography{bibshorter,backjumping.arxiv}

\begin{thebibliography}{}

\bibitem[\protect\citeauthoryear{Bruynooghe}{Bruynooghe}{2004}]{DBLP:journals/tplp/Bruynooghe04}
{\sc Bruynooghe, M.} 2004.
\newblock Enhancing a search algorithm to perform intelligent backtracking.
\newblock {\em Theory Pract. Log. Program.\/}~{\em 4,\/}~3, 371--380.

\bibitem[\protect\citeauthoryear{Bruynooghe and Pereira}{Bruynooghe and
  Pereira}{1984}]{BruynoogheP84}
{\sc Bruynooghe, M.} {\sc and} {\sc Pereira, L.~M.} 1984.
\newblock Deduction revision by intelligent backtracking.
\newblock In {\em Implementations of Prolog}, {J.~A. Campbell}, Ed. Ellis
  Horwood/Halsted Press/Wiley, 194--215.

\bibitem[\protect\citeauthoryear{Deransart, Ed{-}Dbali, and Cervoni}{Deransart
  et~al\mbox{.}}{1996}]{Prolog.standard96}
{\sc Deransart, P.}, {\sc Ed{-}Dbali, A.}, {\sc and} {\sc Cervoni, L.} 1996.
\newblock {\em Prolog - the standard: reference manual}.
\newblock Springer.

\bibitem[\protect\citeauthoryear{Drabent}{Drabent}{2018}]{Drabent.tplp18}
{\sc Drabent, W.} 2018.
\newblock Logic + control: On program construction and verification.
\newblock {\em Theory and Practice of Logic Programming\/}~{\em 18,\/}~1,
  1--29.

\bibitem[\protect\citeauthoryear{Howe and King}{Howe and
  King}{2012}]{howe.king.tcs-shorter}
{\sc Howe, J.~M.} {\sc and} {\sc King, A.} 2012.
\newblock A pearl on {SAT} and {SMT} solving in {Prolog}.
\newblock {\em Theor. Comput. Sci.\/}~{\em 435}, 43--55.

\bibitem[\protect\citeauthoryear{Maher}{Maher}{1988}]{DBLP:books/mk/minker88/Maher88}
{\sc Maher, M.~J.} 1988.
\newblock Equivalences of logic programs.
\newblock In {\em Foundations of Deductive Databases and Logic Programming},
  {J.~Minker}, Ed. Morgan Kaufmann, 627--658.

\bibitem[\protect\citeauthoryear{Robbins, King, and Howe}{Robbins
  et~al\mbox{.}}{2021}]{DBLP:journals/tplp/RobbinsKH21}
{\sc Robbins, E.}, {\sc King, A.}, {\sc and} {\sc Howe, J.~M.} 2021.
\newblock Backjumping is exception handling.
\newblock {\em Theory Pract. Log. Program.\/}~{\em 21,\/}~2, 125--144.

\bibitem[\protect\citeauthoryear{Tarau and Boyer}{Tarau and
  Boyer}{1990}]{DBLP:conf/plilp/TarauB90}
{\sc Tarau, P.} {\sc and} {\sc Boyer, M.} 1990.
\newblock Elementary logic programs.
\newblock In {\em Programming Language Implementation and Logic Programming,
  PLILP'90}, {P.~Deransart} {and} {J.~Maluszynski}, Eds. Lecture Notes in
  Computer Science, vol. 456. Springer, 159--173.

\end{thebibliography}

\end{document}